# Selective injection AlGaN/GaN heterojunction bipolar transistors with patterned regrown base contacts

Chandan Joishi[1], Sheikh Ifatur Rahman[1], Zhanbo Xia[1], Shahadat H. Sohel[1], and Siddharth Rajan[1,2]

[1]*Department of Electrical and Computer Engineering, The Ohio State University, Columbus, OH 43210 USA*
[2]*Department of Electrical and Computer Engineering, The Ohio State University, Columbus, OH 43210 USA*
*Email: joishi.1@osu.edu, rajan@ece.osu.edu*

**Abstract:** We demonstrate graded AlGaN/GaN heterojunction bipolar transistors (HBTs) with selective injection of minority carriers across a p-GaN base and patterned regrown base contacts. The selective injection design regulates minority carrier transport under emitter-base forward bias through a thin base region, while thick and highly doped $p^+$ GaN regrown layers patterned alongside the thin base regions are utilized to lower the base contact resistance. With $SiO_2$ employed as a spacer between the emitter and the $p^+$ regrown layers, the device with an interdigitated emitter/base-contact stripe design displayed a maximum collector current density ($I_C$) of 101 kA/cm$^2$, a maximum current gain ($\beta$) of 70 at $I_C$ ~1 kA/cm$^2$ and ~11 for $I_C$ > 50 kA/cm$^2$. The reported results demonstrate the potential of the selective injection approach to break the long-existing HBT design tradeoff between base resistance and current gain for next-generation radio frequency and mm-Wave applications.

## I. Introduction

GALLIUM nitride heterojunction bipolar transistors (GaN HBTs) combine the intrinsic benefits of wideband gap (3.4 eV) and high saturation velocity (~2x10$^7$ cm/s) of GaN with absence of peak electric field at the device surface/edges [1]. The peak electric field in an HBT is positioned at the base-collector junction in the material's bulk. GaN HBTs, therefore, can operate at near theoretical breakdown electric fields (~3.3 MV/cm) with minimal dispersion from surface states, and better linearity due to a linear relationship between the output current and gain [2]. Moreover, the HBT device miniaturization is done at the epitaxial growth level for the base with Angstrom scale thickness control. Therefore, GaN HBTs provide a solution to mitigate the crucial challenges that limit the high-power handling capability of current GaN HEMTs at radio frequency (RF) and millimeter-wave (mmWave) frequencies [3]. - premature breakdown below material breakdown field (~ 1 MV/cm), current collapse from surface states, breakdown/gain trade-offs from a field-plate design [4, 5].

GaN HBTs demonstrated until now with InGaN, GaN as the base have achieved DC current densities < 30 kA/cm$^2$ with high current gain [6-12], pulsed current densities > 50 kA/cm$^2$ [10, 13], cutoff frequencies ($f_T$) < 8 GHz, and maximum oscillation frequencies ($f_{max}$) < 2 GHz [10] with best reported Johnson's Figure of Merit (JFoM) close to

500 GHz-V, significantly lower than theoretical predictions (> 5 THz-V) [14]. The primary bottleneck to device performance for InGaN base HBTs stems from the high reverse leakage currents across the GaN/InGaN P-N junctions [15]. On the other hand, GaN HBTs with p-GaN base offer low parasitic reverse leakage [9, 16, 17]. These HBTs however, suffer from high base sheet resistivity due to the high activation energy ($E_A$) of magnesium (p-type dopant) on GaN ($E_{A, Mg}$ = 160-220 meV) coupled with low hole mobility (< 20 cm$^2$/Vs) [18]. This leads to current crowding at the emitter edges, primarily caused by a non-negligible voltage drop from the edge to the center of the base active region [2]. While the high base sheet resistance could be lowered by increasing the base layer thickness as well as base doping concentration, both these approaches lead to a decrease in the common emitter current gain and an increase in the base transit delay of the device. As such, approaches to circumvent the trade-off between base resistance and current gain are key to realize GaN HBTs with good performance. Besides, the p-GaN base contact is highly sensitive to plasma etch damage incurred during device processing. This sensitivity poses significant challenges in realizing ohmic base terminals with low contact resistances [18] [19]. The voltage drop at the base contact and the base-emitter access region typically manifests as a high knee and collector offset voltage in the common-emitter output characteristics.

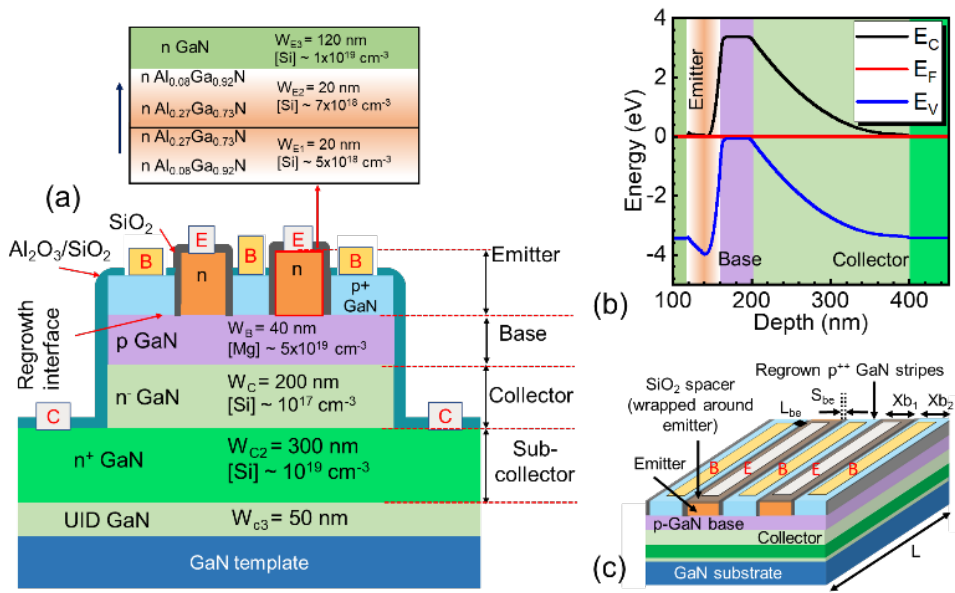

Fig. 1. (a) 2-D schematic of the graded AlGaN/GaN HBT fabricated using a patterned base regrowth process flow, (b) energy band diagram of the HBT, and (c) 3-D schematic of the fabricated device.

One approach to mitigate the base resistance vs device performance trade-off is to tailor the geometry of the p-type base layer itself to minimize current crowding while retaining good device performance. To implement such a design, we propose periodically patterned base stripes of alternating thin p-GaN and thick p$^+$-GaN to realize good base contact behavior via thick, highly doped p$^+$ stripes while channeling minority carrier transport through the thin active base regions. This selective minority carrier injection design reduces base transit delay in the thin base regions while good ohmic behavior from p$^+$ regrowth in conjunction with submicron scaling of the intrinsic base stripes could potentially reduce emitter crowding effects to device characteristics. In this manuscript, we demonstrate the selective injection concept using a patterned base contact regrowth design. The primary advantage of such a design is the presence of a pristine emitter-base junction for current transport which is devoid of unintentional impurities. Existing literature on base contact regrowth designs for GaN HBTs with GaN base are limited to low collector currents (μA scale) and common emitter current gain (β < 5) [1, 20]. We employed interdigitated emitter/base-contact stripe patterns where the unetched emitter sheltered the thin p-GaN base stripe and heavily doped p$^+$ GaN layers were regrown on the etched emitter regions to output a maximum collector current density ($I_C$) of 101 kA/cm$^2$, a maximum current gain (β) of 70 at a current density ~1 kA/cm$^2$ and ~11 for $I_C$ > 50 kA/cm$^2$.

## II. DEVICE GROWTH AND FABRICATION DETAILS

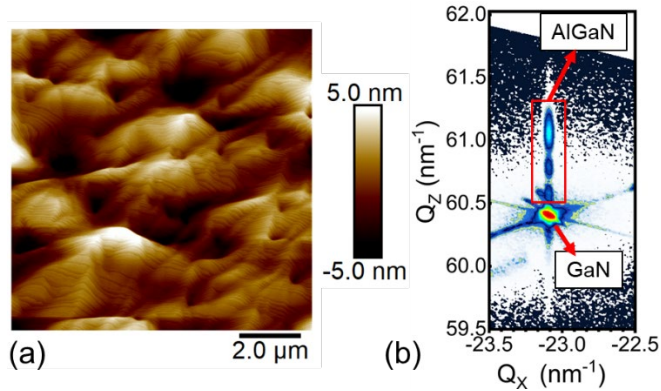

Fig. 2. (a) AFM image of the as-grown epitaxial stack on a 10x10 μm$^2$ surface area ($t_{rms}$ =1.8 nm). (b) XRD reciprocal space mapping data of the as-grown epitaxial stack.

A two-dimensional (2-D) schematic and energy band diagram of the graded AlGaN/GaN HBT are shown in Figures 1(a) and (b). Figure 1(c) shows a 3-D schematic of the interdigitated emitter /p$^+$ regrown base contact stripe geometry. $X_{b1}$ and $X_{b2}$ represent the thin and thick p$^+$ stripe thicknesses respectively, $S_{be}$ is the SiO$_2$ spacer thickness separating the regrown p$^+$ GaN and the emitter layers, and $L_{be}$ denotes the separation between the edge of the emitter layer and the base contact metal. The HBT epitaxial layers were grown on a (0001) c-plane bulk GaN substrate in a VEECO GEN930 plasma assisted molecular beam epitaxy (PAMBE) system equipped with a RF nitrogen (N$_2$) plasma source, and effusion cells of elemental Ga, Si, Al, and Mg. The epitaxial growth was realized in a metal (Ga) rich regime to promote 2-D growth with a nominal growth rate of 4 nm/minute [21-23]. An N$_2$ plasma power of 300 W, a background chamber pressure at ~10$^{-5}$ Torr, and a growth temperature of 750 °C was maintained throughout the process. The growth was initiated with 50 nm GaN UID as a nucleation layer followed by 300 nm of n-doped GaN ([Si] ~ 1x10$^{19}$ cm$^{-3}$) as the sub collector for collector ohmic contact, 200 nm of GaN as the collector layer with a background doping concentration of ~10$^{17}$ cm$^{-3}$, 40 nm of p-GaN ([Mg] ~ 5x10$^{19}$ cm$^{-3}$) as the base, 20 nm of linearly graded AlGaN from 8%→27% ([Si] ~ 5x10$^{18}$ cm$^{-3}$), 20 nm of reverse graded AlGaN from 27%→8% ([Si]~7x10$^{18}$ cm$^{-3}$), and 120 nm of n-GaN ([Si]~10$^{19}$ cm$^{-3}$) for the emitter contact. The as-grown surface resulted in a smooth surface morphology characterized using atomic force microscopy (AFM) measurements with a 2-D step-flow growth, and similar rms roughness as the bulk GaN template (Figure 2 (a)). The layer thicknesses and Al composition were confirmed using High Resolution X-ray diffraction (HR-XRD) (Bruker XRD) measurements and fully strained growth was confirmed using XRD-reciprocal space mapping (RSM) measurements (asymmetric 105 plane) shown in Figure 2(b).

Device fabrication commenced with the formation of interdigitated mask stripes using $SiO_2$ to pattern the regrowth regions for the base contacts. With $SiO_2$ as the hard mask, the emitter layer in the base contact regions was etched using inductively coupled plasma reactive ion etch chamber (ICP-RIE) at a low power etch condition of ~4-5 W RIE and 40 W ICP to minimize plasma etch damage to the p-GaN base surface. Thereafter, the sample was treated with 25% TMAH heated at 85 ºC for 15 mins followed by piranha clean at room temperature for 10 mins [22, 23] to mitigate plasma etch-damage on device sidewalls and the p-GaN surface as well as achieve near vertical sidewalls [24]. 160 nm $SiO_2$ was then deposited on the sample to form a spacer between the emitter and the regrown base layer. The deposition was followed by directional etching of 130 nm of $SiO_2$ with $CF_4/O_2$ based etch-chemistry. The remaining $SiO_2$ was isotopically etched using dilute BOE to avoid etch damage to the p-GaN surface by $CF_4$ plasma and output $S_{be}$ ~100 nm. 60 nm of heavily doped $p^+$ GaN ($[Mg]$~$10^{20}$ $cm^{-3}$) was then regrown on these patterned base contact regions using MBE. The regrowth was preceded by formation of the p-GaN base contacts using Pd/Ni/Au annealed at 450 ºC for 2 minutes in $N_2$ ambient in a rapid thermal annealer. Thereafter, dry etching to expose the emitter and collector contact regions were carried out. Ti/Al/Ni/Au was deposited as the ohmic metal stack in the emitter and collector contact regions. 100 nm PECVD $SiO_2$ followed by 100 nm $Al_2O_3$ using atomic layer deposition (ALD) was then deposited to rest the bond pad layers for emitter and base contacts. The bond pad formation for all the terminals using Ni/Au formed the last fabrication step.

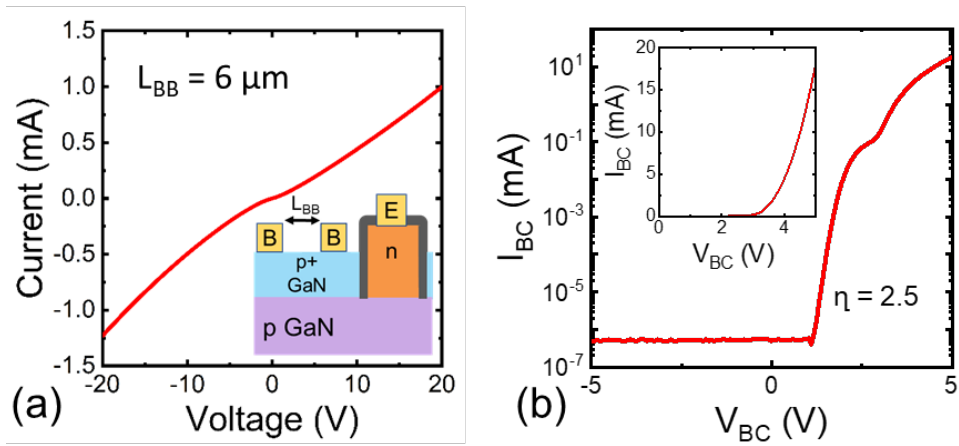

Fig. 3. (a) Two terminal I-V characteristics for adjacent base contacts separation ($L_{BB}$) of 6 μm. (b) Base-collector diode I-V characteristics in semi-logarithmic scale with linear scale in the inset.

III. RESULTS AND DISCUSSIONS

Two terminal I-V characteristics measured between two adjacent base contacts on the regrown $p^+$ GaN layer displayed ohmic behavior as shown in Figure 4(a). A specific contact resistivity ~ $9.3 \times 10^{-3}$ $\Omega.cm^2$ was extracted from TLM measurements on the regrown $p^+$ layer. The as-grown base-collector junction displayed good ON-OFF rectification ratio with ideality factor (η) ~2.5 and turn on voltage ($V_{ON}$) of 3 V for the forward diode characteristics (Figure 3(b)). Low reverse leakage (< 100 μA/$cm^2$) was obtained from the reverse characteristics. The ideality factor (η), which is close to 2, highlights space-charge recombination as the major source of current transport. The Gummel plot, measured at a collector-base voltage ($V_{CB}$) of 0 V to minimize the impact of reverse bias collector-base leakage to the transfer characteristics, is shown in Figure 4(a). The emitter consists of 9 stripes, 1.5x2.6 μm² in dimension with ~ 550 nm separation between an adjacent emitter and a base metal contact stripe. From the Gummel plot, a peak in the common emitter current gain (β) is observed at a collector current density ($I_C$) ~ 1 kA/cm², while β saturates to ~ 11 for $I_C$ > 50 kA/cm². The current density reported here is normalized to the total emitter area. The low hole density in III-Nitride HBTs (estimated to be ~4.8x10¹⁷ cm⁻³) leads to high-injection effects in the base which are well-known as the Webster effect [27]. Similar behaviors have been previously reported for GaN HBTs with low base doping concentration [28]. We estimate that the threshold for the Webster effect is 14 kA/cm² calculated using the equation [33],

$$J_C = \frac{qD_{nB}n(0)}{W_B} \quad (1)$$

where q is the elementary charge, $W_B$ is the base thickness, $D_{nB}$ is the effective electron diffusion coefficient in the base, and n(0) is the injected minority carrier concentration assumed to be equal to the hole density in the base. The calculated current density threshold is in the same regime as observed here. Exact modeling of the Webster-effect as a function of current density is complicated by current crowding in the base (base sheet resistance ~260 kΩ/□). This is qualitatively different from previous reports of "anomalous" current gain [26] which were reported to be an artefact of high base-collector leakage. In the present devices, the leakage current is very low due to the use of low-dislocation density bulk GaN substrates (Figure 3(b)).

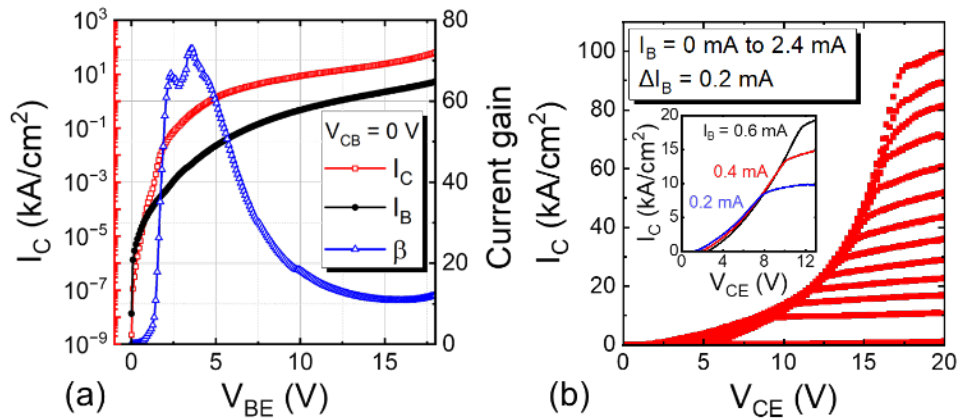

Fig. 4. (a) Gummel plot of the HBT with $V_{CB}$=0 V during the measurement. (b) Common emitter output characteristics of the HBT (output characteristics at low $I_C$ (< 20 kA/cm²) for three different $I_B$ is shown in the inset)

Figure 4(b) shows the common emitter output curves ($I_C$-$V_{CE}$) of the graded AlGaN HBT for $I_B$ ranging from 0 to 2.4 mA in steps of 0.2 mA. A maximum $I_C$ ~ 101 kA/cm² was extracted at $V_{CE}$ = 20 V. The voltage drop across (a) the base contact, (b) the p⁺ GaN regrowth interface, and (c) the lateral base-emitter access region ($L_{be}$) increases $V_{BE}$ needed to sustain a given $I_B$ in the intrinsic device region. This in turn shifts the saturation to active transition of the output curve, i.e., the $V_{CB}$ = 0 line, to the right thereby increasing the knee voltage. Further optimization of the base regrowth is essential to mitigate the high knee voltage of the device. The offset voltage ($V_{CE,offset}$), i.e., the $V_{CE}$ corresponding to which $I_C$ = 0 is observed to be ~ 1 V for $I_C$ ≤ 10 kA/cm² as shown in the inset (Figure 4(b)). When $V_{CE}$ is smaller than

$V_{CE,offset}$, the base-collector junction is forward biased and current flows in the opposite direction out of the collector terminal. With increasing $I_B$ and decreasing current gain, $V_{CE,offset}$ shifts to the right due to an increased collector bias required to mitigate the forward biased parasitic base-collector current.

Figure 5 shows a benchmark plot comparing $I_C$ vs β obtained in this work with previously reported data on GaN HBTs with InGaN and GaN base. While the highest reported $I_C$ using quasi static (pulsed) measurements with an InGaN base is 141 kA/cm$^2$, the reported DC current densities are less than 30 kA/cm$^2$ across all device designs. The present work with $I_C$ = 101 kA/cm$^2$ as such, is the highest output current density reported using DC measurements till date. While the observed current gain is low for reasons explained in the manuscript, improved base epilayer designs with submicron stripe dimensions is expected to further improve the current device metrics.

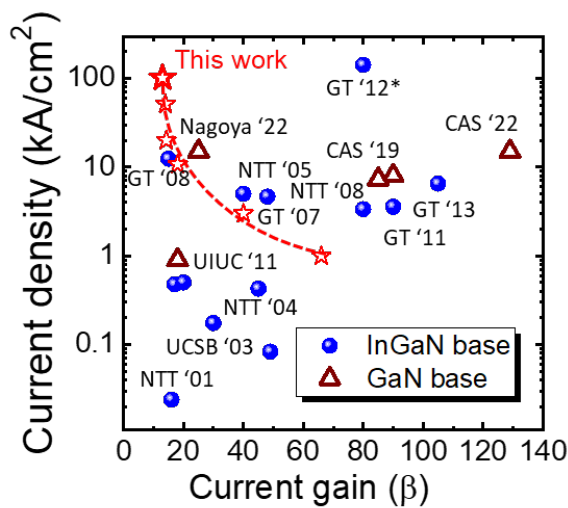

Fig. 5. Benchmarking collector current density ($I_C$) vs common emitter current gain (β) in this work with previously reported data on GaN HBTs with InGaN and GaN base (*data from quasi static I-V measurements with 1 ms pulse, 2% duty cycle) [6-11, 13, 14, 16, 29-32]

## IV.  CONCLUSION

In summary, we demonstrated graded AlGaN/GaN HBT with selectively injected minority carrier transport realized using periodically patterned emitter /p$^+$ base regrown contact stripes to improve device metrics. A combination of low damage plasma etch, wet chemical surface treatments, and base regrowth lead to ohmic base contacts with a pristine n-p-n configuration for intrinsic device operation to output collector current density ~101 kA/cm$^2$, a maximum current gain ~70 at $I_C$ ~ 1 kA/cm$^2$ and ~ 11 for $I_C$ > 50 kA/cm$^2$. While our work highlights the need to minimize base resistance as well as hole doping density in the base, the reported values of current density and gain are very promising and show that continued research on HBTs could enable new flexibility in high-power GaN RF device design.

REFERENCES

[1] L. S. McCarthy *et al.*, "GaN HBT: toward an RF device," *IEEE Transactions on Electron Devices,* vol. 48, no. 3, pp. 543-551, 2001, DOI: 10.1109/16.906449.
[2] W. Liu, *Fundamentals of III-V Devices: HBTs, MESFETs, and HFETs/HEMTs*. Wiley & Sons., 1999.
[3] S. H. Sohel *et al.*, "X-Band Power and Linearity Performance of Compositionally Graded AlGaN Channel Transistors," *IEEE Electron Device Letters,* vol. 39, no. 12, pp. 1884-1887, 2018, DOI: 10.1109/led.2018.2874443.
[4] M. Haziq, S. Falina, A. A. Manaf, H. Kawarada, and M. Syamsul, "Challenges and Opportunities for High-Power and High-Frequency AlGaN/GaN High-Electron-Mobility Transistor (HEMT) Applications: A Review," *Micromachines,* vol. 13, no. 12, p. 2133, 2022.DOI: 10.3390/mi13122133.
[5] M. W. Rahman, N. K. Kalarickal, H. Lee, T. Razzak, and S. Rajan, "Integration of high permittivity BaTiO3 with AlGaN/GaN for near-theoretical breakdown field kV-class transistors," *Applied Physics Letters,* vol. 119, no. 19, 2021, DOI: 10.1063/5.0070665.
[6] L. Zhang *et al.*, "AlGaN/GaN Heterojunction Bipolar Transistors With High Current Gain and Low Specific on-Resistance," *IEEE Transactions on Electron Devices,* vol. 69, no. 12, pp. 6633-6636, 2022, DOI: 10.1109/ted.2022.3217245.
[7] L. Zhang *et al.*, "AlGaN/GaN Heterojunction Bipolar Transistor With Selective-Area Grown Emitter and Improved Base Contact," *IEEE Transactions on Electron Devices,* vol. 66, no. 3, pp. 1197-1201, 2019, DOI: 10.1109/ted.2018.2890207.
[8] S.-C. Shen *et al.*, "GaN/InGaN Heterojunction Bipolar Transistors With $f_T$ > 5 GHz", *IEEE Electron Device Letters,* vol. 32, no. 8, pp. 1065-1067, 2011, DOI: 10.1109/led.2011.2156378.
[9] K. Kumakura and T. Makimoto, "High performance pnp AlGaN∕GaN heterojunction bipolar transistors on GaN substrates," *Applied Physics Letters,* vol. 92, no. 15, 2008, DOI: 10.1063/1.2912502.
[10] R. D. Dupuis *et al.*, "III-N high-power bipolar transistors," *ECS Transactions,* vol. 58, no. 4, pp. 261, 2013, DOI: 10.1149/05804.0261ecst.
[11] B. F. Chu-Kung *et al.*, "Modulation of high current gain (β>49) light-emitting InGaN∕GaN heterojunction bipolar transistors," *Applied Physics Letters,* vol. 91, no. 23, 2007, DOI: 10.1063/1.2821380.
[12] T. Makimoto, Y. Yamauchi, and K. Kumakura, "High-power characteristics of GaN/InGaN double heterojunction bipolar transistors," *Applied Physics Letters,* vol. 84, no. 11, pp. 1964-1966, 2004, DOI: 10.1063/1.1675934.
[13] Y.-C. Lee *et al.*, "GaN/InGaN heterojunction bipolar transistors with ultra-high d.c. power density (>3 MW/cm2)," *physica status solidi (a),* vol. 209, no. 3, pp. 497-500, 2012, DOI: 10.1002/pssa.201100436.
[14] S.-C. Shen *et al.*, "Working toward high-power GaN/InGaN heterojunction bipolar transistors," *Semiconductor Science and Technology,* vol. 28, no. 7, pp. 074025 %@ 0268-1242, 2013, DOI: 10.1088/0268-1242/28/7/074025.
[15] T. Chung *et al.*, "Growth of InGaN HBTS by MOCVD," *Journal of electronic materials,* vol. 35, no. 4, pp. 695-700 %@ 1543-186X, 2006, DOI: 10.1007/s11664-006-0123-z.
[16] X. Huili, P. M. Chavarkar, S. Keller, S. P. DenBaars, and U. K. Mishra, "Very high voltage operation (>330 V) with high current gain of AlGaN/GaN HBTs," *IEEE Electron Device Letters,* vol. 24, no. 3, pp. 141-143, 2003, DOI: 10.1109/led.2003.811400.
[17] B. S. Shelton *et al.*, "AlGaN/GaN heterojunction bipolar transistors grown by metal organic chemical vapour deposition," *Electronics Letters,* vol. 36, no. 1, 2000, DOI: 10.1049/el:20000053.
[18] H. Xing *et al.*, "Progress in gallium nitride-based bipolar transistors," in *Proceedings of the 2001 BIPOLAR/BiCMOS Circuits and Technology Meeting (Cat. No.01CH37212)*, 2-2 Oct. 2001 2001, pp. 125-130, DOI: 10.1109/BIPOL.2001.957873.
[19] J. Chen and W. D. Brewer, "Ohmic Contacts on p-GaN," *Advanced Electronic Materials,* vol. 1, no. 8, 2015, DOI: 10.1002/aelm.201500113.
[20] L. S. McCarthy, P. Kozodoy, M. J. W. Rodwell, S. P. DenBaars and U. K. Mishra, "AlGaN/GaN heterojunction bipolar transistor," in *IEEE Electron Device Letters*, vol. 20, no. 6, pp. 277-279, June 1999, DOI: 10.1109/55.767097.
[21] E. Tarsa, B. Heying, X. Wu, P. Fini, S. DenBaars, and J. Speck, "Homoepitaxial growth of GaN under Ga-stable and N-stable conditions by plasma-assisted molecular beam epitaxy," *Journal of Applied Physics,* vol. 82, no. 11, pp. 5472-5479, 1997, DOI: 10.1063/1.365575.
[22] B. Heying *et al.*, "Optimization of the surface morphologies and electron mobilities in GaN grown by plasma-assisted molecular beam epitaxy," *Applied Physics Letters,* vol. 77, no. 18, pp. 2885-2887, 2000, DOI: 10.1063/1.1322370.
[23] B. Heying, R. Averbeck, L. F. Chen, E. Haus, H. Riechert, and J. S. Speck, "Control of GaN surface morphologies using plasma-assisted molecular beam epitaxy," *Journal of Applied Physics,* vol. 88, no. 4, pp. 1855-1860, 2000, DOI: 10.1063/1.1305830.
[24] Y. Zhang *et al.*, "Trench formation and corner rounding in vertical GaN power devices," *Applied Physics Letters,* vol. 110, no. 19, 2017, DOI: 10.1063/1.4983558.
[25] Pao-Chuan Shih, Zachary Engel, Habib Ahmad, William Alan Doolittle, Tomás Palacios, "Wet-based digital etching on GaN and AlGaN", *Applied Physics Letters*. 10 January 2022; 120 (2): 022101. DOI: 10.1063/5.0074443.
[26] H. Xing, D. Jena, M. J. W. Rodwell, and U. K. Mishra, "Explanation of anomalously high current gain observed in GaN based bipolar transistors," *IEEE Electron Device Letters,* vol. 24, no. 1, pp. 4-6, 2003, DOI: 10.1109/led.2002.807023.
[27] W. M. Webster, "On the Variation of Junction-Transistor Current-Amplification Factor with Emitter Current," *Proceedings of the IRE,* vol. 42, no. 6, pp. 914-920, 1954, DOI: 10.1109/JRPROC.1954.274751.
[28] Kazuhide Kumakura, Toshiki Makimoto, "High-voltage operation with high current gain of pnp AlGaN∕GaN heterojunction bipolar transistors with thin n-type GaN base", *Appl. Phys. Lett*. 10 January 2005; 86 (2): 023506. DOI: 10.1063/1.1851608
[29] K. Kumakura, T. Makimoto, and N. Kobayashi, "Common-emitter current–voltage characteristics of a Pnp AlGaN/GaN heterojunction bipolar transistor with a low-resistance base layer," *Applied Physics Letters,* vol. 80, no. 20, pp. 3841-3843, 2002, DOI: 10.1063/1.1480102.
[30] T. Makimoto, K. Kumakura, and N. Kobayashi, "High current gains obtained by InGaN/GaN double heterojunction bipolar transistors with p-InGaN base," *Applied Physics Letters,* vol. 79, no. 3, pp. 380-381, 2001, DOI: 10.1063/1.1387261.
[31] T. Kumabe *et al.*, ""Regrowth-free" fabrication of high-current-gain AlGaN/GaN heterojunction bipolar transistor with N-p-n configuration," *Applied Physics Express,* vol. 15, no. 4, 2022, DOI: 10.35848/1882-0786/ac6197.
[32] S. Shyh-Chiang *et al.*, "Surface Leakage in GaN/InGaN Double Heterojunction Bipolar Transistors," *IEEE Electron Device Letters,* vol. 30, no. 11, pp. 1119-1121, 2009, DOI: 10.1109/led.2009.2030373.
[33] W. Liu. "*Handbook of III-V heterojunction bipolar transistors"*, John Wiley & Sons., pp. 160, 1998